# Die Wellenmaschine

- Grundlagen der Wellenausbreitung, Dispersion, Reflexion, Simulation -

Tilman Küpper*

*Hochschule für angewandte Wissenschaften München, tilman.kuepper@hm.edu
(6. März 2015)

**Kurzfassung**

Wellenmaschinen werden seit vielen Jahren im Physikunterricht eingesetzt, um die Ausbreitung, Überlagerung und Reflexion von Wellen anschaulich vorzuführen. Physikalische Größen wie Schwingungsfrequenz, Wellenlänge und Phasengeschwindigkeit werden üblicherweise experimentell ermittelt. Der vorliegende Artikel zeigt, dass – ausgehend von den technischen Daten der Wellenmaschine – auch eine Berechnung dieser charakteristischen Größen möglich ist. Dämpfungseinrichtungen zur Verhinderung der Reflexion einlaufender Wellen werden ebenfalls betrachtet.

Ein im wahrsten Sinne des Wortes schönes Projekt für den Informatikunterricht ist die Simulation von Wellenmaschinen am Computer mit einer animierten Darstellung der Simulationsergebnisse. Einige Details hierzu sind gegen Ende dieses Artikels zusammengestellt.

**1. Einleitung**

Koppelt man eine größere Zahl beweglicher Pendel mit Federn aneinander, so erhält man eine Anordnung, auf der die Ausbreitung von Transversalwellen anschaulich vorgeführt werden kann. Solche „Wellenmaschinen" (Abbildung 1) werden bereits seit vielen Jahren erfolgreich im Physikunterricht eingesetzt [8, 9, 10] und auch im Rahmen von Informatik-Projektarbeiten am Computer simuliert.

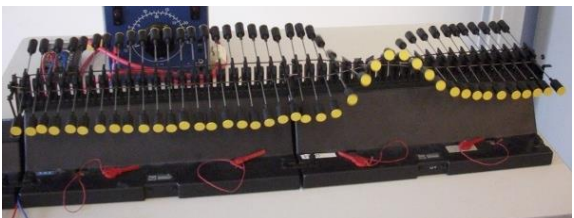

**Abb.1:** Wellenmaschine[1]

Bei genauer Beobachtung der Wellenausbreitung auf diesen Maschinen ergeben sich weiterführende Fragen, auf einige davon wird im vorliegenden Artikel näher eingegangen:

- Wellenlänge und Phasengeschwindigkeit werden im Unterricht in Versuchen gemessen. Ist es auch möglich, ausgehend von den technischen Daten der Wellenmaschine, diese Werte zu berechnen?

- Im Gegensatz beispielsweise zur Ausbreitung elektromagnetischer Wellen im Vakuum zeigt sich eine leichte Abhängigkeit der Phasengeschwindigkeit von der Schwingungsfrequenz. Wodurch kommt diese Abhängigkeit zustande? Kann dieser Zusammenhang mathematisch beschrieben werden?

- Viele Wellenmaschinen sind mit Dämpfungseinrichtungen ausgestattet, mit denen einlaufende Wellen absorbiert und Reflexionen verhindert werden sollen. Deren Einstellung ist freilich nicht ganz einfach, oft stören „Reflexionsreste" die nachfolgenden Beobachtungen. Warum ist die korrekte Einstellung der Dämpfungseinrichtungen so schwierig? Kann die Größe der Reflexionsreste abgeschätzt werden?

**2. Ausbreitung von Wellen**

Die folgenden Überlegungen erfolgen auf Grundlage eines vereinfachten Wellenmaschinenmodells, einer linearen Kette von Massekugeln, wobei jeweils zwei benachbarte Kugeln über eine Feder miteinander verbunden sind (Abbildung 2).

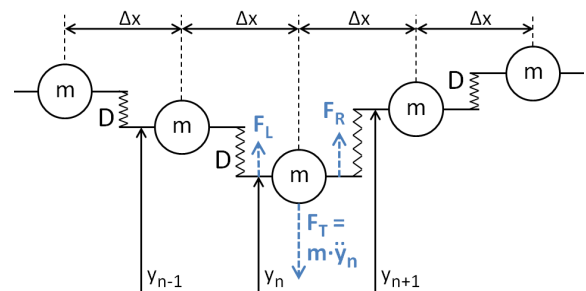

**Abb.2:** Vereinfachtes Wellenmaschinenmodell

---

[1] Foto: K.-H. Meyberg (http://gfs.khmeyberg.de).

Bei der Ausbreitung von Wellen entlang dieser Kette sind zum einen die Federkräfte $F_L$ und $F_R$, zum anderen die Trägheitskräfte $F_T$, welche durch die Beschleunigung der Kugeln verursacht werden, von Bedeutung. Für die n-te Kugel gilt folgende Kräftebilanz:

$$m\ddot{y}_n = F_T = F_L + F_R$$
$$m\ddot{y}_n = (y_{n-1} - y_n)D + (y_{n+1} - y_n)D$$
$$\frac{m}{D}\ddot{y}_n = y_{n-1} + y_{n+1} - 2y_n \qquad \{1\}$$

Wenn eine sinusförmige Welle von links nach rechts über die Wellenmaschine wandert, schwingen alle Kugeln mit derselben Frequenz und Amplitude, lediglich die Nullphasenwinkel sind unterschiedlich:

$$y_{n-1}(t) = \hat{Y}\sin(\omega t + \varphi_w)$$
$$y_n(t) = \hat{Y}\sin(\omega t)$$
$$y_{n+1}(t) = \hat{Y}\sin(\omega t - \varphi_w) \qquad \{2\}$$

Der Übergang zu komplexen Drehzeigern[2] erleichtert die nachfolgenden Berechnungen:

$$\underline{y}_{n-1}(t) = \hat{Y}e^{i\omega t}e^{+i\varphi_w}$$
$$\underline{y}_n(t) = \hat{Y}e^{i\omega t}$$
$$\underline{y}_{n+1}(t) = \hat{Y}e^{i\omega t}e^{-i\varphi_w} \qquad \{3\}$$
$$\underline{\dot{y}}_n(t) = i\omega\hat{Y}e^{i\omega t}$$
$$\underline{\ddot{y}}_n(t) = -\omega^2\hat{Y}e^{i\omega t} \qquad \{4\}$$

Einsetzen in Gleichung {1} führt zu:

$$(e^{+i\varphi_w} + e^{-i\varphi_w}) - 2 = -\frac{m}{D}\omega^2$$
$$(2\cos\varphi_w) - 2 = -\frac{m}{D}\omega^2 \qquad \{5\}$$

Bei bekannter Federkonstante $D$, Kugelmasse $m$ und Frequenz $\omega$ kann damit die Phasenverschiebung $\varphi_w$ zwischen den Schwingungen zweier benachbarter Kugeln angegeben werden:

$$1 - \cos\varphi_w = \frac{m\omega^2}{2D} \qquad \{6\}$$

Diese Gleichung wird mit dem Halbwinkelsatz $1 - \cos\varphi_w = 2\sin^2\frac{\varphi_w}{2}$ weiter vereinfacht und nach $\varphi_w$ aufgelöst:

$$2\sin^2\frac{\varphi_w}{2} = \frac{m\omega^2}{2D}$$
$$\varphi_w = 2\arcsin\sqrt{\frac{m\omega^2}{4D}} \qquad \{7\}$$

### 2.1. Wellenlänge und Phasengeschwindigkeit

Eine vollständige Schwingungsperiode umfasst $N_\lambda$ benachbarte Kugeln:

$$N_\lambda = \frac{2\pi}{\varphi_w} \qquad \{8\}$$

Setzt man $\varphi_w$ (im Bogenmaß) aus Gleichung {7} ein, kann schließlich die Wellenlänge $\lambda$ berechnet werden:

$$\lambda = N_\lambda \Delta x = \frac{2\pi}{\varphi_w}\Delta x = \frac{\pi\Delta x}{\arcsin\sqrt{\frac{m\omega^2}{4D}}} \qquad \{9\}$$

---

[2] Die Schwingung $y(t) = \hat{Y}\sin(\omega t + \varphi)$ im Zeitbereich wird durch den Drehzeiger $\underline{y}(t) = \hat{Y}e^{i(\omega t + \varphi)}$ in der komplexen Ebene dargestellt [2]. Es gilt der Zusammenhang $y(t) = \text{Im}\{\underline{y}(t)\}$.

Für die Phasengeschwindigkeit $c$ folgt entsprechend:

$$c = \frac{\omega}{2\pi}\lambda = \frac{\omega\Delta x}{2\arcsin\sqrt{\frac{m\omega^2}{4D}}} \qquad \{10\}$$

Die Phasengeschwindigkeit ist demnach abhängig von der Frequenz $\omega = 2\pi f$ (sog. Dispersion [1, 6]). Abbildung 3 zeigt die Phasengeschwindigkeit $c$ in Abhängigkeit von $f$, als Kugelmasse wurde $m$ = 0,01 kg, als Federkonstante $D$ = 1 Nm$^{-1}$ und als Kugelabstand $\Delta x$ = 0,025 m angenommen. Oberhalb einer Grenzfrequenz $f_0$ = 3,18 Hz ist keine Wellenausbreitung möglich: Das Argument der arcsin-Funktion in den Gleichungen {7}, {9} und {10} wäre dann größer als 1, die Wellenlänge $\lambda$ würde den doppelten Kugelabstand unterschreiten.

Statt der Abhängigkeit $c = f(\omega)$ wird in der Literatur häufig die sog. „Dispersionsrelation" $\omega = f(k)$ angegeben (mit der „Wellenzahl" k) [1]:

$$k = \frac{2\pi}{\lambda} = \frac{2}{\Delta x}\arcsin\sqrt{\frac{m\omega^2}{4D}}$$
$$\sin\frac{k\Delta x}{2} = \sqrt{\frac{m\omega^2}{4D}}$$
$$\omega = \sqrt{\frac{4D}{m}}\sin\frac{k\Delta x}{2} \qquad \{11\}$$

### 2.2. Ausbreitung niederfrequenter Wellen

Abbildung 3 zeigt, dass die Phasengeschwindigkeit für niedrige Frequenzen nahezu konstant ist. Erst wenn bei höheren Frequenzen die Wellenlänge $\lambda$ nicht mehr groß gegenüber dem Kugelabstand $\Delta x$ ist, „bemerkt"[3] die Welle etwas von der diskreten Struktur der Wellenmaschine, die Phasengeschwindigkeit sinkt.

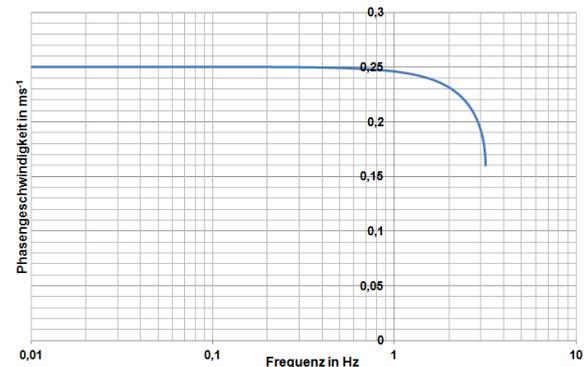

**Abb.3:** Phasengeschwindigkeit und Frequenz

Die Wellenlänge $\lambda_0$ bei *niedrigen Frequenzen* kann ausgehend von Gleichung 9 abgeschätzt werden:

$$\lambda = \frac{\pi\Delta x}{\arcsin\sqrt{\frac{m\omega^2}{4D}}}$$
$$\lambda_0 \approx \frac{\pi\Delta x}{\sqrt{\frac{m\omega^2}{4D}}} = \frac{2\pi\Delta x}{\omega}\sqrt{\frac{D}{m}} \qquad \{12\}$$

---

[3] vergl. [3], Seite 4

Für die Phasengeschwindigkeit $c_0$ bei niedrigen Frequenzen gilt entsprechend:

$$c_0 = \frac{\omega}{2\pi}\lambda_0 \approx \Delta x \sqrt{\frac{D}{m}} \quad \{13\}$$

## 3. Reflexion am Ende der Wellenmaschine

Läuft eine Welle von links nach rechts auf das Ende der Wellenmaschine zu, wird sie dort im Allgemeinen reflektiert. Eine fest eingespannte letzte Kugel am Ende der Wellenmaschine führt zu einer Reflexion am festen Ende (ein Wellenberg läuft als „Wellental" zurück), eine frei bewegliche letzte Kugel führt zu einer Reflexion am offenen Ende (ein Wellenberg läuft als Wellenberg zurück).

Tritt bei der letzten Kugel am Ende der Wellenmaschine eine geschwindigkeitsabhängige Reibungskraft der Größe $F_{Reib} = \alpha \dot{y}$ auf, führt somit ein Reibungskoeffizient $\alpha \to \infty$ zur Reflexion am festen und ein Reibungskoeffizient $\alpha = 0$ zur Reflexion am offenen Ende.

Gibt es aber nun einen bestimmten Reibungskoeffizienten $\alpha_0$, bei dem die Welle am Ende der Wellenmaschine ohne jegliche Reflexion vollständig absorbiert wird?

Angenommen, die n-te Kugel bildet das rechte Ende der Wellenmaschine. Dann muss die Reibungskraft $F_{Reib,n}$ an der n-ten Kugel jederzeit exakt derjenigen Federkraft entsprechen, die zwischen der n-ten und (n+1)-ten Kugel einer „unendlich langen" Wellenmaschine auftreten würde:

$$F_{Reib,n} = \alpha_0 \dot{y}_n \stackrel{!}{=} D(y_n - y_{n+1})$$
$$\alpha_0 i\omega e^{i\omega t} = D(1 - e^{-i\varphi_w})e^{i\omega t} \quad \{14\}$$

Mit $1 - e^{-i\varphi_w} = i\sqrt{2 - 2\cos\varphi_w} \cdot e^{-i\frac{\varphi_w}{2}}$ (Kosinussatz) und nach Einsetzen von Gleichung {6} folgt:

$$i\omega\alpha_0 e^{i\omega t} = i\omega\sqrt{Dm} \cdot e^{-i\frac{\varphi_w}{2}} e^{i\omega t} \quad \{15\}$$

Damit kann $\alpha_0$ angegeben werden:

$$\alpha_0 = \sqrt{Dm} \cdot e^{-i\frac{\varphi_w}{2}} \quad \{16\}$$

Es existiert also *kein reeller* Reibungskoeffizient $\alpha_0$, der einen reflexionsfreien Abschluss der Wellenmaschine ermöglicht.

Für einen solchen reflexionsfreien Abschluss müsste die Reibungskraft nach Gleichung {16} mit einer Phasenverschiebung von $-\varphi_w/2$ am Ende der Wellenmaschine angreifen, was allerdings technisch nicht realisierbar ist. Denn die Reibungskraft ergibt sich stets aus der der *aktuellen* Kugelgeschwindigkeit.

Lediglich bei kleinen Winkeln $\varphi_w$ – also bei großen Wellenlängen $\lambda \gg 2\Delta x$ bzw. entsprechend niedrigen Schwingungsfrequenzen $f$ – ist $e^{-i\frac{\varphi_w}{2}} \approx 1$ und es gilt näherungsweise:

$$\alpha_0 \approx \sqrt{Dm} \quad \{17\}$$

### 3.1. Reflexionsfreier Abschluss in der Praxis

Da ein vollständig reflexionsfreier Abschluss der Wellenmaschine durch Reibung an der letzten Kugel offensichtlich nicht möglich ist, soll nun betrachtet werden, inwiefern Reflexionen durch den technisch realisierbaren reellen Reibungskoeffizienten aus Gleichung {17} tatsächlich unterdrückt werden und so zumindest ein „praktisch reflexionsfreier Abschluss" durch eine entsprechende Dämpfungseinrichtung realisiert werden kann.

Für die folgenden Betrachtungen sei erneut vorausgesetzt, dass die n-te Kugel das rechte Ende der Wellenmaschine darstellt. Dann wirken auf diese n-te Kugel drei unterschiedliche Kräfte:

- die Federkraft zwischen der (n-1)-ten und n-ten Kugel,
- die Trägheitskraft aufgrund der Beschleunigung der n-ten Kugel,
- die Reibungskraft mit dem Reibungskoeffizienten $\alpha_0$.

Für die n-te Kugel gilt damit folgendes Kräftegleichgewicht:

$$D(y_{n-1} - y_n) - m\ddot{y}_n - \alpha_0 \dot{y}_n = 0$$
$$\text{mit } \alpha_0 = \sqrt{Dm}$$
$$y_{n-1} - y_n - \frac{m}{D}\ddot{y}_n - \sqrt{\frac{m}{D}}\dot{y}_n = 0 \quad \{18\}$$

Ohne Beschränkung der Allgemeinheit sei angenommen, dass die n-te Kugel eine sinusförmige Schwingung der Amplitude 1 mit verschwindendem Nullphasenwinkel ausführt:

$$\underline{y}_n = e^{i\omega t}$$
$$\underline{\dot{y}}_n = i\omega e^{i\omega t}$$
$$\underline{\ddot{y}}_n = -\omega^2 e^{i\omega t} \quad \{19\}$$

Aus dem Kräftegleichgewicht (Gleichung {18}) folgen damit Amplitude $Z$ und Nullphasenwinkel $\varphi_Z$ der (n-1)-ten Kugel:

$$\underline{Z}e^{i\omega t} - e^{i\omega t} + \frac{m}{D}\omega^2 e^{i\omega t} - \sqrt{\frac{m}{D}}j\omega e^{i\omega t} = 0$$
$$\text{mit } \underline{Z} = Ze^{i\varphi_Z}$$
$$\underline{Z} = 1 - \frac{m\omega^2}{D} + i\sqrt{\frac{m\omega^2}{D}} \quad \{20\}$$

Mit der Abkürzung $\xi = \sqrt{\frac{m\omega^2}{D}}$ ergibt sich eine etwas kompaktere Darstellung:

$$\underline{Z} = 1 - \xi^2 + i\xi \quad \{21\}$$

Die Wellenmaschine ist wie bereits dargestellt nicht vollständig reflexionsfrei abgeschlossen, es handelt sich bei der Bewegung von (n-1)-ter und n-ter Kugel also um die Überlagerung einer hinlaufenden und einer rücklaufenden (reflektierten) Welle. Mit den komplexen Drehzeigern $\underline{H} = He^{i\varphi_H}$ und $\underline{R} = Re^{i\varphi_R}$ zur Darstellung der Amplituden und Nullphasenwinkel dieser beiden Wellenanteile nehmen die Be-wegungsgleichungen der beiden Kugeln folgende Form an (bei den hin- und rücklaufenden Wellenanteilen tritt wiederum die bereits bekannte Phasenverschiebung $\varphi_w$ zwischen den benachbarten letzten beiden Kugeln auf):

$$\underline{y}_n = \underline{H}e^{i\omega t} + \underline{R}e^{i\omega t}$$
$$\underline{y}_{n-1} = \underline{H}e^{i\omega t}e^{i\varphi_w} + \underline{R}e^{i\omega t}e^{-i\varphi_w} \qquad \{22\}$$

Gleichsetzen mit Gleichung {19} bzw. Gleichung {21} ergibt:

$$1 \stackrel{!}{=} \underline{H} + \underline{R}$$
$$Z = 1 - \xi^2 + i\xi \stackrel{!}{=} \underline{H}e^{i\varphi_w} + \underline{R}e^{-i\varphi_w} \qquad \{23\}$$

Durch Einsetzen der oberen in die untere Gleichung können zunächst $\underline{R}$ und anschließend $\underline{H}$ bestimmt werden:

$$1 - \xi^2 + i\xi = (1 - \underline{R})e^{i\varphi_w} + \underline{R}e^{-i\varphi_w} \qquad \{24\}$$
$$\underline{R} = -\frac{1 - \xi^2 + i\xi - e^{i\varphi_w}}{2i \sin \varphi_w}$$
$$\underline{H} = 1 + \frac{1 - \xi^2 + i\xi - e^{i\varphi_w}}{2i \sin \varphi_w} \qquad \{25\}$$

Der Reflexionsfaktor $r = |\underline{R}|/|\underline{H}|$, also das Amplitudenverhältnis zwischen rück- und hinlaufenden Wellenanteilen, kann nun angegeben werden. Es ergibt sich eine etwas „unhandliche" Formel, die freilich am Computer leicht auszuwerten ist:

$$r = \left| \frac{m\omega^2 - 2i\sqrt{m\omega^2 D} + i\sqrt{m\omega^2(4D - m\omega^2)}}{-4D \sin\left[2 \arcsin \sqrt{\frac{m\omega^2}{4D}}\right] - im\omega^2 - 2\sqrt{m\omega^2 D} + \sqrt{m\omega^2(4D - m\omega^2)}} \right| \qquad \{26\}$$

Abbildung 4 zeigt den Verlauf des Reflexionsfaktors $r$ in Abhängigkeit von der Frequenz $f$, für die technischen Daten der Wellenmaschine wurden dieselben Werte wie in Abbildung 3 angenommen, der Reibungskoeffizient ist $\alpha_0 = \sqrt{Dm}$.

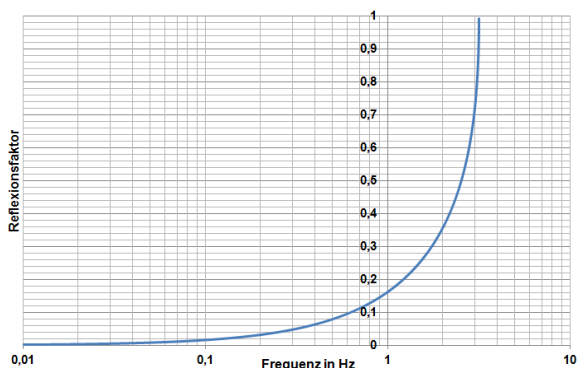

**Abb.4:** Reflexionsfaktor und Frequenz

Für einen nahezu reflexionsfreien Abschluss der Wellenmaschine sollte $r$ natürlich möglichst klein sein, im Idealfall gleich null. Sofern die Schwingungsfrequenz mehr als eine Dekade unterhalb der Grenzfrequenz der Wellenmaschine bleibt (hier $f_0$ = 3,18 Hz – siehe Abschnitt 2.1), kann in der Tat von einem „praktisch reflexionsfreien" Abschluss der Wellenmaschine ausgegangen werden. So ergibt sich bei $f$ = 0,318 Hz gerade einmal ein Reflexionsfaktor von $r \approx 5\%$.

Abbildung 4 macht aber auch deutlich, warum höherfrequente Störungen nur langsam von der Wellenmaschine verschwinden: Knapp unterhalb der Grenzfrequenz $f_0$ = 3,18 Hz liegt der Reflexionsfaktor nahezu bei eins. Solche Störungen werden also selbst bei eingeschalteter Dämpfungseinrichtung kaum absorbiert.

**4. Simulation am Computer**

Ein – auch im wörtlichen Sinn – schönes Projekt für den Informatikunterricht ist die Simulation von Wellenmaschinen am Computer, wozu natürlich die animierte Darstellung der Simulationsergebnisse gehört [11].

Die Simulation basiert auf Gleichung {1}. Stellt man diese Kräftebilanz für alle n Kugeln der Wellenmaschine auf, führt das zunächst zu einem System von n Differentialgleichungen zweiter Ordnung. Zur einfacheren Programmierung werden diese in ein System von 2·n Differentialgleichungen erster Ordnung umgewandelt, dazu werden neue Zustandsvariablen $u_i$ eingeführt:

$u_1 := y_1$ (1. Kugel, Auslenkung)
$u_2 := \dot{y}_1$ (1. Kugel, Geschwindigkeit)
$u_3 := y_2$ (2. Kugel, Auslenkung)
$u_4 := \dot{y}_2$ (2. Kugel, Geschwindigkeit)
...
$u_{2n-1} := y_n$ (n-te Kugel, Auslenkung)
$u_{2n} := \dot{y}_{n-1}$ (n-te Kugel, Geschwindigkeit)
$\qquad \{27\}$

Es ergibt sich folgendes System erster Ordnung:

...

2. Kugel $\begin{cases} \dot{u}_3 = u_4 \\ \dot{u}_4 = \dfrac{D}{m}(u_1 + u_5 - 2u_3) \end{cases}$

3. Kugel $\begin{cases} \dot{u}_5 = u_6 \\ \dot{u}_6 = \dfrac{D}{m}(u_3 + u_7 - 2u_5) \end{cases}$

... $\qquad \{28\}$

Die erste und die letzte (n-te) Kugel der Wellenmaschine werden gesondert behandelt. Am linken Rand erfolgt zum Beispiel eine sinusförmige Anregung, am rechten Rand führt geschwindigkeitsproportionale Reibung mit dem Reibungskoeffizienten α entweder zu Reflexion oder zu mehr oder weniger ausgeprägter Absorption:

1. Kugel $\begin{cases} \dot{u}_1 = u_2 \\ \dot{u}_2 = \dfrac{D}{m}[\hat{Y}\sin(\omega t) + u_3 - 2u_1] \end{cases}$

n-te Kugel $\begin{cases} \dot{u}_{2n-1} = u_{2n} \\ \dot{u}_{2n} = \dfrac{D}{m}(u_{2n-3} + u_{2n-1}) - \dfrac{\alpha u_{2n}}{m} \end{cases}$

$\qquad \{29\}$

Die WellenmaschineStandardverfahren zur numerischen Integration solcher Differentialgleichungssysteme erster Ordnung wie das explizite Euler-Verfahren oder Einschrittverfahren nach Runge-Kutta sind in der einschlägigen Literatur beschrieben[4]. Für Programmiersprachen wie C oder Java sind sie als fertige Module kostenlos verfügbar, bei MATLAB sind sie sogar fest integriert.

In Anhang A ist ein einfaches MATLAB-Skript zur Simulation einer Wellenmaschine abgedruckt. Die Wellenmaschine wird am linken Rand sinusförmig angeregt, rechts ist sie mit $\alpha = \sqrt{Dm}$ „praktisch reflexionsfrei" abgeschlossen. Abbildung 5 zeigt ein Bildschirmfoto des Programmablaufs.

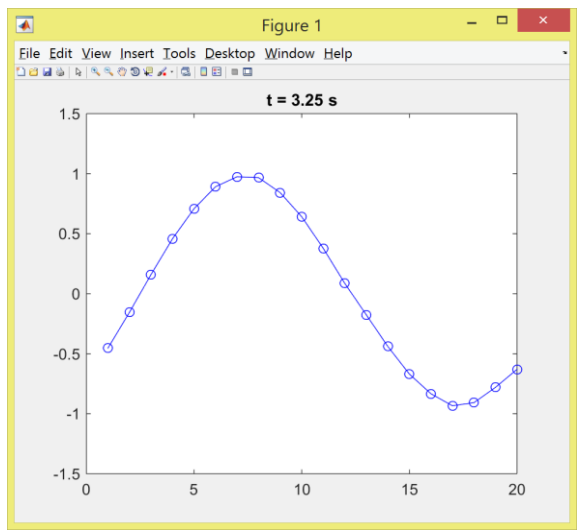

**Abb.5:** Simulation mit MATLAB

### 5. Schlussbemerkungen

Im Mittelpunkt des vorliegenden Artikels steht die Berechnung charakteristischer Größen der Wellenausbreitung auf Wellenmaschinen. Es zeigt sich, dass die Phasengeschwindigkeit bei Vergrößerung der Schwingungsfrequenz absinkt (Dispersion).[5]

Weiterführende Betrachtungen zur Reflexion und Absorption von Wellen zeigen, warum selbst bei präziser Einstellung von Dämpfungseinrichtungen am Ende der Wellenmaschine gewisse „Reflexionsreste" nicht verhindert werden können.

Zur weiteren Lektüre seien insbesondere die Veröffentlichungen von J. Shive empfohlen, einem Halbleiterphysiker, der in den 1950er-Jahren eine erste Version der Wellenmaschine für den Unterricht an Schulen und Colleges in den USA entwickelt hat. In seinem Buch „Similarities in Wave Behavior" [4] betrachtet er neben den Grundlagen der Wellenausbreitung, Überlagerung und Reflexion von Wellen das Übertragungsverhalten und Impedanzanpassung von Wellenleitern. In „Characteristics of Electrons in Solids" [5] nutzt er eine leicht modifizierte Wellenmaschine zur Erläuterung des Energiebändermodells von Festkörpern.

### 6. Quellenverzeichnis

[1] Meschede, Dieter (Hrsg.) (2010): Gerthsen Physik, 24. Auflage; Springer Berlin/Heidelberg/ New York, Kapitel 4 (deformierbare Körper, Schwingungen und Wellen), S. 139-209

[2] Arens, Tilo; Hettlich, Frank et al. (2008): Mathematik, 1. Auflage; Spektrum Akademischer Verlag, Heidelberg, Kapitel 5 (komplexe Zahlen), S. 121-146

[3] Hübel, Horst (2006): Zur Quantenphysik der linearen Kette; http://www.forphys.de/Website/qm/linkette.pdf (Stand: 04/2012)

[4] Shive, John (1964): Similarities in Wave Behavior, Teacher's Edition; Bell Telephone Laboratories

[5] Shive, John (1960): Characteristics of Electrons in Solids. In: IRE Transactions on Education, vol. 3, issue 4, pp. 106-110

[6] Burgel, B. A. (1967): Dispersion, Reflection, and Eigenfrequencies on the Wave Machine. In: American Journal of Physics, vol. 35, issue 10, pp. 913-915

[7] Di Renzone, Simone; Frati, Serena; Montalbano, Vera (2012): Disciplinary Knots and Learning Problems in Waves Physics; 12th International Symposium Frontiers of Fundamental Physics, Udine; eprint arXiv: 1201.3008

[8] Staatsinstitut für Schulqualität und Bildungsforschung München (2007): Atome, Wellen, Quanten; ISB-Handreichung Gymnasium, Physik, Kapitel 2 (Wellen und Quanten), S. 71-87

[9] Phywe Systeme GmbH, Göttingen: Betriebsanleitung Wellenmaschine 11211.00; http://www.phywe.de (Stand: 04/2012)

[10] Leybold Didactic GmbH, Hürth: Gebrauchsanweisung Wellenmaschine 40120-24; http://ld-didactic.de (Stand: 04/2012)

[11] Landesbildungsserver Baden-Württemberg des Landesinstituts für Schulentwicklung (LS), Stuttgart: Virtuelle Wellenmaschine, Simulation einer Wellenmaschine aus 97 gekoppelten Oszillatoren, http://www.schule-bw.de/unterricht/faecher/physik/online_material/wellen/mech_wellen/lk5 (Stand: 07/2014)---

[4] Siehe zum Beispiel [2], Seite 430ff.

[5] Das diesem Artikel zugrunde liegende vereinfachte Wellenmaschinenmodell beschreibt die Ausbreitung von Transversalwellen. Vergleichbare Überlegungen zur Ausbreitung von Torsionswellen wurden 1967 von B. A. Burgel veröffentlicht [6].

**Anhang A - Programmbeispiel in MATLAB**

```matlab
% Simulation einer Wellenmaschine, Hauptprogramm
function wellensim()
    global D f m n reib

    m = 0.01;   D = 1;      % Kugelmasse, Federkonstante
    n = 20;     f = 0.5;    % Kugelanzahl, Anregungsfreq.
    reib = sqrt(D*m);       % Reibungskoeffizient

    % Zu Beginn sind alle Kugeln in Ruhe
    anf = zeros(1, 2*n);

    % DGL-System für den Zeitraum t = 0...100s lösen
    [T,Y] = ode23(@dgl_system, 0:0.025:100, anf);

    % Animation der Simulationsergebnisse
    clc
    disp('*** Beenden der Animation mit Strg+C ***');
    for i=1:length(T)
        plot(Y(i, 1:2:2*n), 'bo-');
        ylim([-1.5 1.5]);
        title(sprintf('t = %.2f s', T(i)));
        drawnow; pause(0.1);
    end
end

% DGL-System der Wellenmaschine
function dudt = dgl_system(t, u)
    global D f m n reib

    % Erste Kugel, Gleichung {29a}
    dudt(1) = u(2);
    dudt(2) = D/m * (sin(2*pi*f*t) + u(3) - 2*u(1));

    % n-te Kugel, Gleichung {29b}
    dudt(2*n-1) = u(2*n);
    dudt(2*n) = D/m * (u(2*n-3) - u(2*n-1)) - reib*u(2*n)/m;

    % Alle anderen Kugeln, Gleichung {28}
    for i=2:n-1
        dudt(2*i-1) = u(2*i);
        dudt(2*i) = D/m * (u(2*i-3) + u(2*i+1) - 2*u(2*i-1));
    end
    dudt = dudt';
end
```